\documentclass[aps,twocolumn,showpacs]{revtex4}
\usepackage{graphicx}
\usepackage{amsfonts}
\begin{document}

\title{A simple model for some unusual properties of
martensitic transformation}
\author{ S. Sreekala$^1$,
          Rajeev Ahluwalia$^{2}$,
         and G. Ananthakrishna$^{(1,3)}$
     \footnote{Electronic Mail: garani@mrc.iisc.ernet.in}
       }
\affiliation{$^1$ Materials Research Centre, Indian Institute of Science, Bangalore-560012, India\\
$^{2}$ Theoretical Division, Los Alamos National Laboratory,\\
Los Alamos, New Mexico 87545\\
$^{3}$ Centre for Condensed Matter Theory, Indian Institute of
Science, Bangalore-560012, India\\
}

\begin{abstract}
We report a detailed numerical investigation of a recently
introduced two dimensional model for square-to-rectangle
martensitic transformation that explains several unusual features
of the martensitic transformation. This model includes inertial
effects, dissipation, long-range interaction between the
transformed domains and  an inhomogeneous stress field to describe
the effect of lattice defects which serves as nucleation centers.
Both single-site nucleation and multi-site nucleation has been
studied for single quench situation and thermal cycling. The final
stage morphologies of single-site nucleation and multi-site
nucleation bear considerable similarity suggesting that the
initial distribution of the defects is not important. Thermal
cycling using  continuous cooling and heating simulations show the
existence of hysteresis in the transformation. More importantly,
the rate of energy dissipated occurs in the forms of bursts with
power law statistics for their amplitudes and durations which
explains the results of acoustic emission signals observed in
experiments. When the system is cycled repeatedly in a restricted
domain of temperatures, the dissipated bursts of energy are
repetitive, a feature  observed in experiments. The associated
morphology shows a complete reversal of the martensite domains
thus throwing light on the mechanism underlying the shape memory
effect. The model also exhibits tweed like patterns.
\end{abstract}
\maketitle

\section{INTRODUCTION}
\subsection{Motivation}

Martensitic transformation often exhibits unusual features that
are not expected of a first
order transition. One such effect is the well documented
pretransitional effect observed as the system approaches the
martensitic transformation temperature \cite{Muto,Oshima, Robertson, Kartha95}.
 The associated enhanced
levels of fluctuation observed in several measurable quantities (
such as the anomalous scattering and tweed structure)  has been
recorded in a number of different systems
\cite{Kartha95,Muto,Robertson}. This feature, however, is the
signature of   critical fluctuations in a second order transition.
As another example, consider an observation that relates to
acoustic emission (AE), a feature that  usually accompanies
martensite transformation. Recently, in experiments on Cu-Al-Zn
single crystals,  Vives {\it et al} reported that the
distributions of the amplitudes of the AE signals and their
durations obey power law statistics when the samples were
subjected to slow thermal cycling \cite{Vives94,Vives} both during
cooling and heating runs. The statistics were reported to be
robust over a range of cooling and heating rates. Again, power law
statistics  imply scale free nature of the underlying process.
However, in this case, as the AE signals are accumulated during
thermal cycling, there is no tuning as in second order transition.
Thus, it is actually reminiscent of self-organized criticality
introduced by Bak {\it et al } \cite{Bak}.   Another unusual and
interesting property of the AE signals reported is the high degree
of reproducibility and statistical correlation in time when the
system is subjected to repeated thermal cycling over a restricted
range of temperatures \cite{Lovey}. The near repetitive nature of
AE signals during successive cycles have been shown to be
correlated with the growth and shrinkage of martensite domains.
Thus, exploring this correlated behavior should help us to
understand the shape memory effect as well. Moreover, at a
conceptual level, the nature of correlation of the repetitive AE
signals is significantly different from the power law nature of
the statistics observed during full thermal cycling mentioned
above. Thus, it would be desirable to capture the seemingly
conflicting properties using a single model. The purpose of the
paper is to report a detailed study of a recently introduced two
dimensional phenomenological model that is shown to explain both
the power law statistics and correlated behavior of AE signals
\cite{Rajeev,Kala}. As we shall show, the model also captures the
precursor effect.

\subsection{ Background}

The martensitic transformations are technologically important
class of first order, solid-solid,  diffusionless structural phase
transformations. Unlike other phase transformations where
diffusion disperses the neighbouring atoms, here, neighbouring
atoms in the parent phase remain so in the product phase also
\cite{Jdphys}. However, the lattice gets distorted due to
spontaneous displacement of the atoms ( from their positions in
the parent lattice) accompanying the discontinuous change in the
shape and symmetry of the unit cell. This creates long-range
strain fields which in turn strongly depend on the  relative
positions and orientations of the martensitic plates. (Typically,
the martensite morphology consists of thin plate-like domains with
twinned structure that are oriented along the elastically
favorable habit plane directions.) Thus,  the transformation path
depends on continuously evolving configuration dependent long-
range strain fields that eventually leaves the system in a
metastable state. Thus, given a fixed quench, and hence a fixed
amount of drive, the amount of transformed phase is fixed and
further undercooling would be required to increase the product
phase. For the same reason, the transformation occurs over a wide
range of temperatures. On cooling, the transformation starts at a
temperature $M_s$, called the martensite start temperature and is
completed at a temperature $M_f$ (martensite finish temperature).
In the reverse  heating cycle, the transformation is initiated at
a temperature $A_s$ (austenite start temperature ) and ends at a
temperature $A_f$ (austenite finish temperature), which in general
can be much higher than the martensitic start temperature $M_s$.
The related shape memory effect usually accompanies the
transformation. The above features also imply that thermal
fluctuations have little role in the transformation kinetics. Such
martensites are called athermal. It must be mentioned that
martensitic transformations can be induced by applying external
stress. The combined use of temperature and stress have found many
practical applications \cite{xxx}.

In athermal martensites,  the martensitic phase nucleates from
isolated regions in the crystal  which are usually defects like
dislocations or grain boundaries \cite{ferra, Brook, Khacha}.
Thus, quenched disorder plays an important role in the initial
kinetics of athermal martensites. Another interesting feature of
these transformations is the autocatalytic  cooperative nature of
nucleation \cite{ferra}. It is observed that a martensite domain
that has nucleated from a specific site in the crystal, triggers
the nucleation of other domains in the vicinity. Such a correlated
nucleation was observed by Ferraglio and Mukherjee \cite{ferra}
who studied heterogeneous nucleation in Au-Cd alloys.

A large body of theoretical work has accumulated over the years
\cite{Jdphys,Roit,Khacha}. ( See \cite{Olson} for a recent summary
of the status of theory.) Recent approaches use a continuum field
theoretic models with strain or displacement fields as order
parameters \cite{Jacobs,Krum,Cao,Kartha}. Within this framework,
Cao {\it et al }\cite{Cao} have dealt with the problem of
heterogeneous nucleation at localized defect sites by simulating
the influence of defects through an inhomogeneous stress field.
  Recently, Wang and Khachaturyan \cite{Wang} have investigated the
dynamics of improper martensitic transformations based on a time
dependent Ginzburg-Landau (TGDL) approach. This approach  has also
been extended to proper martenstic transformation \cite{Arte}.

Here, we recall a few models relevant for the present work. The
tweed structure has been explained as arising due to inhomogeneous
distribution of the components of the martensitic alloy, for
instance, Pd in Fe \cite{Kartha} using a strain based model. On the
other hand, the power law behaviour of avalanches occurring in
first order transitions has been modelled by using disorder based
Ising models \cite{Sethna}. In this case, the power law
statistics of avalanches arises in the presence of a critical
amount of disorder. Since quenched-in-disorder (defects) plays an
essential role in the nucleation process of martensitic
transformation, it appears that these kinds of models
\cite{Sethna} may be relevant to martensitic transformation.
However,  by subjecting the system to repeated hysteresis cycles,
Vives {\it et al}, have verified that the system evolves towards a
critical state independent of the initial treatment of the alloy
suggesting a dynamical evolution of the system towards such a
state. According to these authors \cite{Vives94}, the interpretation is that in
real martensites, although there are quenched-in defects, the
disorder that is responsible for the power law is actually
generated during the transformation itself. Indeed, this
feature with lack of tuning is reminiscent of self-organized
criticality, a word coined for slowly driven spatially extended
systems evolving to a critical state where the statistics of
avalanches obey a power law \cite{Bak}. Since the introduction of
this concept, there are large number of reports of physical systems
exhibiting SOC like features, for example, earthquakes
\cite{Rich}, acoustic emission from volcanic rocks \cite{Diod},
stress drops during the Portevin Le-Chatelier effect \cite{Anan99}
and biological evolution \cite{Bak93}, to name a few.

Recently, a simple two dimensional phenomenological model has been shown
to capture  the power law statistics  under thermal cycling
\cite{Rajeev}. This model attempts to incorporate the essential
features  of systems evolving to SOC state, namely, slow driving,
threshold dynamics, appropriate relaxational mode without any
recourse to tuning any relevant parameter.  Surprisingly, this
model also captures the correlated nature of the AE signals when
the system is cycled in a restricted range of temperatures
\cite{Kala}, with the associated growth
and decay of the martensite domains.

The purpose of this paper is to present the results of an extensive
numerical simulations on this two dimensional model describing
square-to-rectangle martensitic transformation \cite{Rajeev,Kala}.
We discuss the results of a single-defect quench, multi-defect
quench, thermal cycling over broad range of temperatures, the
power law statistics arising in both in single and multi-defect
cases, the correlated nature of AE signals and the associated
shape memory effect,  and finally  the tweed like structure.

The organization of this paper is as follows. In section II, we
describe our model starting from a free-energy functional and
derive an equation of motion  for a strain order parameter. The
model has  three free parameter one of which is the temperature.
Section III contains the results of a detailed numerical
simulations  carried out for various conditions of quench
parameters. In section IV, we consider thermal cycling for single
and multi-site cases both of which exhibit thermal hystersis. By
calculating the energy dissipated during the transformation, we
show that the distribution of the amplitudes and durations of the
energy bursts during thermal cycling obey power law statistics as
in experiments \cite{Vives}. Section V deals with the correlated
behavior of AE signals and its correspondence with shape memory
features. Section VI, deals with the tweed structure. We end the
paper with some observations on the model.

\section{ THE MODEL }

The basic idea of the model is to include all the important
features of athermal martentites such as  inertial effects, long-
range interaction and dissipation. The inertial effect is included
by accounting for  finite propagation time in a manner similar to
that considered by Bales and Gooding \cite{Bales, Reid}. These
authors have demonstrated that inertial effects prohibit the
growth of martensite as a single variant in the presence of
dissipation (in one dimension). Instead, the martensite grows as
an alternating arrangement of the two variants. The reason for
including dissipation stems from the recognition that the
parent-product interface moves at near velocity of sound as
suggested by the emission of AE signals. Associated with this
movement, there is a dissipation which tends to relax the system
towards local equilibrium. We include this through a Rayleigh
dissipation functional \cite{Landau}. ( Note that conventional
phase transformation take place at sufficiently slow pace thereby
providing  adequate time for quasi-steady state conditions to be
attained in a short time.) We also include heterogenous nucleation
at defect sites by including an appropriate strain energy
\cite{Cao}.

In order to capture these seemingly different types of features
within the scope of a single model, it is desirable to keep the
model as simple as possible. To this end we first consider a {\it
2d} square-to-rectangle transition \cite{Krum,Jacobs}. However,
the free energy  in general depends on all the three strain
components defined by
\begin{eqnarray}
e_{1}&=&{{(\eta_{11}+\eta_{22})}\over {\sqrt{2}}},\nonumber\\
e_{2}&=&{{(\eta_{11}-\eta_{22})}\over {\sqrt{2}}},\nonumber\\
e_{3}&=&\eta_{12}=\eta_{21},
\end{eqnarray}
with
\begin{equation}
\eta_{ij}={{1}\over{2}}({{\partial u_{i}}\over{\partial x_{j}}}
+{{\partial u_{j}}\over{\partial x_{i}}})
\end{equation}
referring to the components of the linearized strain tensor and $u_i$'s  are
the displacement fields in the direction $i$ ( $i=x,y$). The
components $e_{1}, e_2$ and $e_3$ are the bulk dilational strain,
deviatoric strain and shear strain respectively. The
simplification we make is to assume that it is adequate to
consider the deviatoric strains to be the principal order
parameter. This is a reasonable assumption considering the fact
that  volume changes are usually small. Henceforth, we denote the
deviatoric strain by $\epsilon(\vec{r})$ and define
\begin{equation}
\epsilon(\vec{r})=(\frac{\partial u_{x}(\vec{r})}{\partial x}
-\frac{\partial u_y(\vec{r})}{\partial y})/\sqrt 2 = \epsilon_x(\vec r) - \epsilon_y(\vec r),
\end{equation}
where $u_{x}$ and $u_{y}$ are respectively the displacement fields in the $x$ and $y$
directions.\\

The free-energy functional of our system with the order-parameter
$\epsilon$ is written as
\begin{equation}
F\{\epsilon(\vec{r})\}=
{F_{L}}\{\epsilon(\vec{r})\}+
{F_{LR}}\{\epsilon(\vec{r})\},
\end{equation}
where $F_{L}$ is a local free-energy functional and $F_{LR}$ is a
nonlocal long-range term that describes transformation induced
strain-strain interaction. In a scaled form, we write the local
free-energy $F_L$ as
\begin{equation}
F_L=\int d\vec{r}\bigg[f_l(\epsilon(\vec{r}))+
{D\over{2}}{{({\nabla}\epsilon(\vec{r}))^{2}}}
-
\sigma(\vec{r})\epsilon(\vec{r})\bigg].
\end{equation}
where $D$ and $\sigma$ are in a scaled form. The latter represents
the stress field due to localized defects in the crystal. (In a real
crystal, lattice defects like dislocations and grain boundaries
act as sources of stress concentration.) In the above  equation,
$f_l(\epsilon(\vec{r}))$ is the usual Landau polynomial for a
first order transition given by
\begin{equation}
f_l(\epsilon(\vec{r}))={{\tau}\over{2}}{\epsilon(\vec{r})}^{2}
-{\epsilon(\vec{r})}^{4}+
{{1}\over{2}}{\epsilon(\vec{r})}^{6}.
\end{equation}
Here, $\tau = (T - T_c)/(T_0 - T_c)$ is the scaled temperature.
$T_0$ is the first-order transition temperature at which the free
energy for the product and parent phases are equal, and $T_c$ is
the  temperature below which there are only two degenerate global
minima $\epsilon = \pm \epsilon_{M}$. The stress field
$\sigma(\vec r)$ in Eqn. 5 modifies the free-energy $f_l$ in such
a way that the austenitic phase is locally unstable leading to the
nucleation of the product phase.

The physical cause of the long-range interaction is the coherency in
strain at the parent-product as well as the product-product interfaces.
 An effective long-
range interaction between the deviatoric  strains  of the
transformed domains has been shown to result from the elimination
of the other strain components, $e_1$ and $e_3$, using St. Venant
compatibility constraint \cite{Kartha,Shenoy}.  Apart from
determining the dependence on $\vec {r}$, one important feature is
that such a kernel picks out the correct habit plane directions
which in the present case are $[11]$ and $[1\bar1]$. As one of our
objectives is {\it to keep the model as simple as possible}, we
have resorted to introducing a long-range kernel in a
phenomenological way retaining the feature that allows the growth
of martensitic domains along the habit plane directions
\cite{Remark}.(In the present case also, it is possible to use
this approach to obtain the appropriate kernel). Wang and
Khachaturyan \cite{Wang} have shown that the interface can be
described by accounting for coherency strains at the
parent-product interface by including  symmetry allowed fourth
order anisotropic long-range interaction in the free energy, ie, a
term which is invariant under $\epsilon \rightarrow - \epsilon$.
We define the long-range interaction by
\begin{equation}
F_{LR}\{\epsilon\}=-{{1}\over{2}}\int \int d\vec{r}d\vec{r'}G(\vec{r}-\vec{r'})
\epsilon^{2}(\vec{r})
\epsilon^{2}(\vec{r'}).
\end{equation}
As much as in the physical situation, in our model also, long-
range interaction plays an important role  in describing the
growth of martensite domains.  The kernel $G(\vec{r}-\vec{r'})$ is
best defined by considering the Fourier representation of the
long-range term given by
\begin{equation}
F_{LR}\{\epsilon\}={{1}\over{2}}\int d\vec{k} B\bigg({{\vec{k}}\over{k}}\bigg)
{\{ {\epsilon^{2}}(\vec{r})\}}_{k}
{\{ {\epsilon^{2}}(\vec{r})\}}_{k^{*}},
\end{equation}
where $\{ {\epsilon^{2}}(\vec{r})\}_{k}$ is the Fourier transform of
$\epsilon^{2}(\vec{r})$ defined as
\begin{equation}
\{ {\epsilon^{2}}(\vec{r})\}_{k}=\int {{d\vec{r}}\over{(2\pi)^{2}}}
{{\epsilon}^{2}}(\vec{r})exp(i\vec{k}\cdot\vec{r}).
\end{equation}
The quantity ${\{ {\epsilon^{2}}(\vec{r})\}}_{k^{*}}$ is the
complex conjugate of  ${\{ {\epsilon^{2}}(\vec{r})\}}_{k}$. The
direction dependent kernel $B({{\vec{k}}/{k}})$ contains
information about the crystallographic details of the crystal and
defines the habit plane. Apart from the favorable directions of
growth of  the product phase along $[11]$ and $[1\bar{1}]$, the
free-energy barriers should be large along the $[10]$ and $[01]$
directions. These features are well captured by the simple kernel
\begin{equation}
B\bigg({{\vec{k}}\over{k}}\bigg)= -\beta \theta(k-\Lambda){{\hat{k}}^{2}}_{x}
{{\hat{k}}^{2}}_{y},
\end{equation}
where ${\hat{k}}_{x}$ and ${\hat{k}}_{y}$ are the unit vectors in
$x$ and $y$ directions (The step function $\theta(k-\Lambda)$ has
been introduced to impose a cutoff on the range of the long-range
interaction.) The constant $\beta$ is the strength of the
interaction. This kernel incorporates the effect of the interface
in a simple way  as the cost of growth progressively increases
with the transformation  in directions where the kernel is
positive which not only aids growth along the habit plane
directions but also limits the growth of domains transverse to it.
We stress that this is only a simple choice and is not unique.
Other kernels with similar orientation dependence will give
similar results\cite{Remark}.  The real space picture of
$B({\vec{k}}/k)$ is similar to the long-range interaction of
Kartha {\it et al} \cite{Kartha95}.

Even though we have taken the deviatoric strain as the order
parameter, basic variables are the displacement fields.  Thus, we
start with the Lagrangian $L=T-F$,  where $F$ is the total
free-energy and $T$ is the kinetic energy associated with the
system. The kinetic energy is given by
\begin{equation}
T=\int d\vec{r}\rho\bigg[\bigg({{\partial u_{x}(\vec{r},t)}\over{\partial t}}
\bigg)^{2}+
\bigg({{\partial u_{y}(\vec{r},t)}\over{\partial t}}\bigg)^{2}\bigg ].
\end{equation}
Here $\rho$ is the mass density. As mentioned in the introduction,
since the parent-product interface moves rapidly, it
is associated  with a dissipation which we have represented by the
Rayleigh dissipative functional \cite{Landau}. Further, since
deviatoric strains are the dominant ones, dissipative functional
is written entirely in terms of $\epsilon(\vec{r})$,
\begin{equation}
R={{1}\over{2}}\gamma\int d\vec{r}{\big({{\partial }\over{\partial
t}}\epsilon (\vec{r},t)\big)}^{2}.
\end{equation}
(Here we have assumed that the bulk and shear strains equilibrate
 rapidly and hence do not contribute to the dissipation function.)
Now consider the possibility of relating the above term to
acoustic energy. As far as we know there are  no attempts to
capture the essential features of acoustic signals in the context of
martensitic transformation. To model the AE signals, we recall
that the mechanism of generation of the AE signals is generally
attributed to the sudden release of the stored strain energy. One
area where there has been some efforts to model AE signals is in plasticity.
In this case, the production of AE signals is attributed to
the abrupt motion of the dislocations. Consequently, the energy of
AE signals, $E_{ae}(r)$ is taken to be proportional $\dot
\epsilon^2(r)$, where $\dot \epsilon$ is the local plastic strain
rate \cite{Weiss}. However, in general there is spatial
inhomogeneity. Then the leading contribution to total energy $E_{ae}\propto
\int (\nabla \dot \epsilon)^2 d^3r$.  This clearly has the same form as
the Rayleigh dissipation functional \cite{Landau} arising from the
rapid movement of a localized region. Thus, while comparing
results of the statistics of AE bursts, we need to simply compute
$R(t)$.

We derive the equations of motion for $\epsilon$ using the
equations of motion for the displacement fields given by
\begin{equation}
{{d}\over{dt}}\bigg({{\delta L}\over{\delta {\dot{u}_i}}}\bigg)-
{{\delta L}\over{\delta u_{i}}}=-{ {\delta R}\over {\delta \dot{u}_i}}, \, \, i = x, y.
\end{equation}

\noindent
Using the above equation, after computing the functional derivatives, we get (see Appendix for details)
\begin{equation}
\rho{{{\partial}^{2} }\over{{\partial t}^{2}}}\epsilon(\vec{r},t)=
{\nabla}^2\bigg[{{\delta F}\over{\delta \epsilon(\vec{r},t)}}
+\gamma {{\partial }\over{\partial t}}\epsilon(\vec{r},t)\bigg],
\end{equation} which after scaling out $\rho$ and $D$ can be
written in the form (in terms of rescaled space and time
variables)
\begin{eqnarray}
\nonumber {{{\partial}^{2} }\over{{\partial
t}^{2}}}\epsilon(\vec{r},t)& = & {\nabla}^2\bigg[{{\partial
f(\vec{r},t)}\over{\partial \epsilon(\vec{r},t)}}- \sigma(\vec{r})
- {\nabla}^2\epsilon(\vec{r},t)+
\gamma {{\partial }\over{\partial t}}\epsilon(\vec{r},t) \\
& + &2\epsilon(\vec{r},t)\int d\vec{k}B(\vec
{k}/k)\{\epsilon^{2}(\vec{k},t)\}_{k}
 e^{i \vec {k} .\vec {r}} \bigg],
\end{eqnarray}
Here, both $\beta$ and $\gamma$ are to be taken as rescaled
parameters. The structure of Eqn. (15) is similar to that derived
in \cite{Bales} for 1-d except for the  long-range term. In the
next section, we use the above equation to study the morphological
evolution during martensitic transformations.

\section {NUMERICAL SIMULATIONS}

We now describe the results of our numerical simulation for the
morphological features.  We discretize Eqn.(15) on a $N \times N$
grid using the Euler's scheme with periodic boundary conditions.
The mesh size of the grid is $\Delta x=1$ and the smallest time
step $\Delta t=0.002$. Most results reported here correspond to $
N = 128,256$. However, wherever necessary, we have carried out
simulations for higher $N$.  A psuedo-spectral technique is
employed to compute the long-range term \cite{Desai}. In this
method, we compute the discrete Fourier transforms of
$\epsilon^2(\vec{r},t)$ and $G(\vec{r})$, take the product and
then calculate the inverse Fourier transform. In all simulations
reported in the paper, the cutoff $\Lambda$ in the long-range
expression defined in  Eq.(10) is chosen to be 0.2. The
inhomogeneous stress field $\sigma(\vec{r})$ is appropriately
chosen to describe the defect configuration (see below). We
consider two  situations corresponding to the nucleation at a
single- defect site and  at several defect  sites.

\subsection {Nucleation at a Single-Defect Site}
We begin with the study of nucleation and growth of domains under
a single quench, starting from the austenite phase to the
martensite phase.  In general, the stress field due to several
defects located at $\vec{r}_j$ can be described  by
\begin{equation}
\sigma(\vec{r})={\sum_j^{j_{max}}}\sigma_0(\vec{r_j}
)exp\bigg({{-|\vec{r}-\vec{r_j}|^{2}}\over{\zeta_j^{2}}}\bigg),
\label{defect}
\end{equation}
where $\sigma_0 (\vec {r}_j)$ is the magnitude of the stress field at
sites $\vec{r}_j$ which are randomly chosen defect sites,
$j_{max}$ is the total number of defect sites, and $\zeta_j$  is the
width of the field.

We first consider a single isotropic defect with its core located
at the center of the system. Although, a single spherically
symmetric defect is a rather artificial system, it is useful to
clarify the physics of nucleation and growth in martensitic
systems. For a single-defect case, $\vec{r}_j = \vec{r}_0$. We
choose $\zeta=1$ and $\sigma_0=0.3$. With this value of
$\sigma_0$, the system becomes locally unstable at the core
$\vec{r}_0$. ( The threshold value of  $\sigma_0$ which makes the
austenite phase unstable at that point is first determined and a
slightly larger value is used. ) The parameters chosen for the
simulations are $\beta=50$ and $\gamma =4$. At $t=0$, we start
with $\epsilon(\vec{r},0)$ distributed in the interval
$[-0.005,0.005]$ representing the austenite phase and
simultaneously 'turn on' the stress field $\sigma(\vec{r})$ as we
quench the system to $\tau = -2.0$.

 Figure 1 shows the nucleation and growth of the martensite
domains from the defect core at various instants of time. (Grey
regions represent the austenite phase with $\epsilon =0$, black
and white regions  represent the two variants of martensite
characterized by $\epsilon =\pm \epsilon_{eq}$.  ) We note here
that the magnitude of   strain in the two domains $\vert
\epsilon_{eq}\vert $  are larger than  $\epsilon_{M}$ ( obtained
from Eq. (6)) due to the contribution arising from the long-range
term.  In a short time after quench ($t \sim 7.5$), we observe the
emergence of a small nucleus with  $\epsilon = \epsilon_{eq}$. In
addition, we also see the emergence of domains of the other
variant $(\epsilon=-\epsilon_{eq})$ adjacent to the nucleus in the
$[11]$ and $[1\bar{1}]$ directions. (We remark here that the
mechanism of twinning in this model is the same as that discussed
in \cite{Bales,Reid}, i.e, kinetic energy minimization in the
presence  of dissipation.) The structure further develops into
twinned arrays, propagating along $[11]$ and $[1\bar{1}]$
directions, as can be clearly seen in the snapshot at time
$t=16.25$. This snap shot also reveals an interesting feature
namely the creation of  nuclei ( $\epsilon = \epsilon_{eq}$)
located close to the $+x$ and $-y$ directions, located at a finite
distance from the propagating arrays. As the twinned fronts
propagate, several additional nuclei are created  at finite
distances from the original propagating fronts as can seen from
the snap shot at $t = 18.5$. This can be attributed to the
accumulation of long -range stress fields at these sites. These
new nuclei give birth to secondary fronts that also propagate
along $[11]$ and $[1\bar{1}]$. These observations in our
simulation are in accordance with the collective nucleation
mechanism discussed in \cite{Wang} and the experiments by
Ferraglio and Mukherjee \cite{ferra}. The propagation of these new
secondary fronts continues till they 'collide' with the
pre-existing martensite domains and stop ($t=20$). The
morphological evolution eventually stops beyond $t=50$.

Figure  2, shows the corresponding evolution of the area fraction
$\phi$ of the martensitic phase ($\circ$). The area fraction is
computed by counting the number of points on the grid for which
$|\epsilon(\vec{r},t)| > 0.5$. The transformation is seen to start
around $t \sim 15$. The fraction increases sharply till about $t
\sim 20$, beyond which it saturates to a value close to $0.31$.
The spurt in the growth between $t=15$ and 20 roughly coincides
with the creation of the first set of additional nuclei.
\begin{figure}
\includegraphics[height=4.5cm]{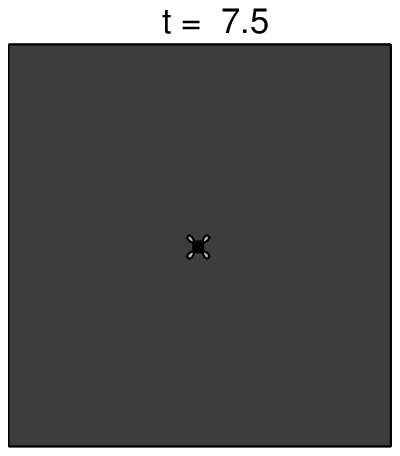}
\includegraphics[height=4.5cm]{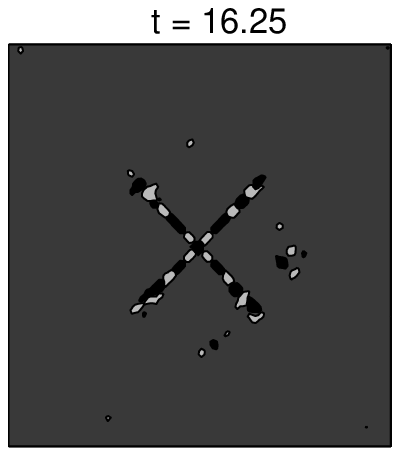}
\includegraphics[height=4.5cm]{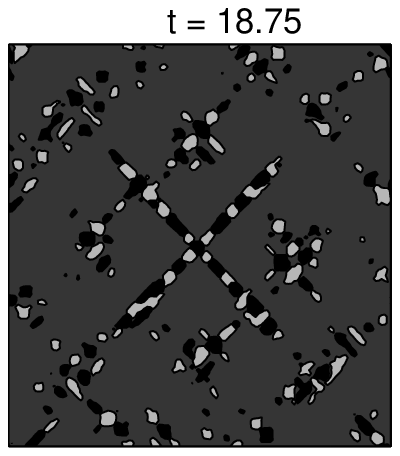}
\includegraphics[height=4.5cm]{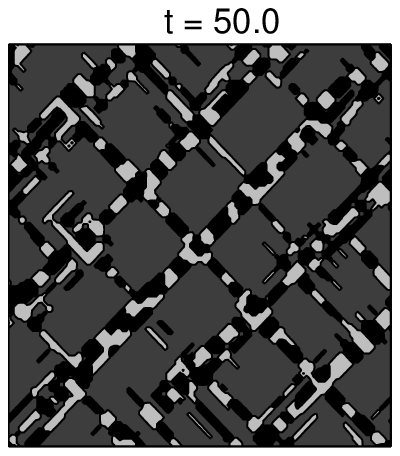}
\caption{Morphological evolution for nucleation at a single-
defect with $\beta = 50$, $\gamma = 4$ and $\tau = -2.0$. Grey
cells correspond to the austenite
 phase and the black and the white to the two martensite variants.}
 \label{fig1}
 \end{figure}

\begin{figure}
\includegraphics[height=4cm]{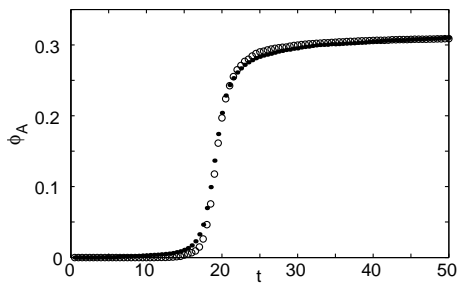}
  \caption{Plot of the transformed area fraction $\phi_A $ with respect to
  time for N=128, $\beta=50$, $\gamma=4$,and $\tau=-2.0$. $\circ$ correspond
to single-site case and $\bullet $ corresponds to the multi-site
defect case. }
  \label{fig2}
\end{figure}

 \begin{figure}
 \includegraphics[height=4.5cm]{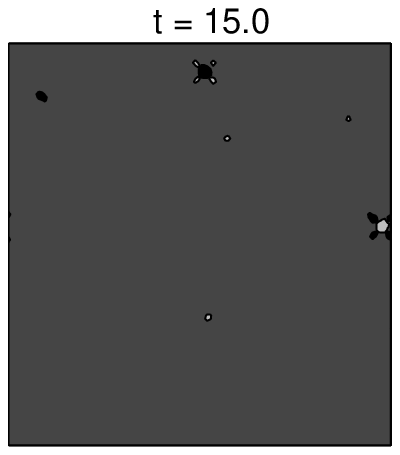}
 \includegraphics[height=4.5cm]{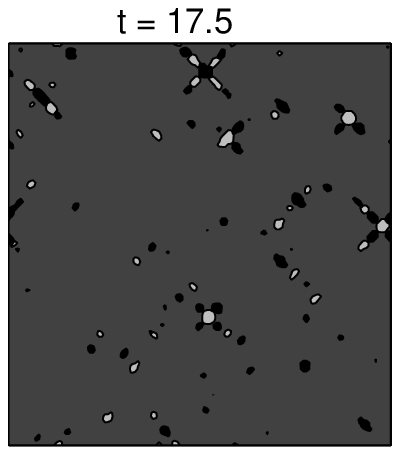}
 \includegraphics[height=4.5cm]{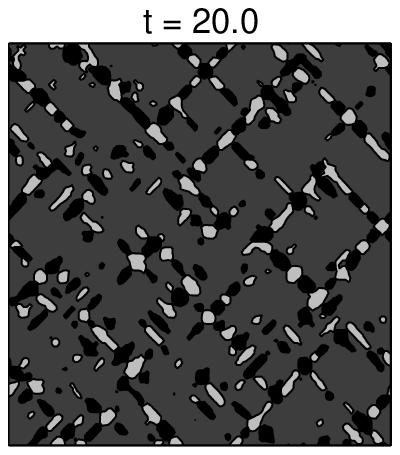}
 \includegraphics[height=4.5cm]{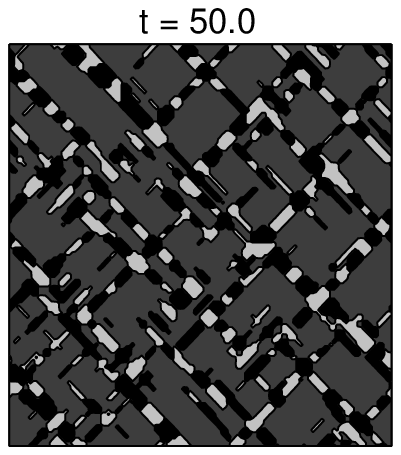}
 \caption{Morphological evolution for multi-defect case  with defect
 density $0.1 \%$ for $\beta=50$, $\gamma=4$ and $\tau=-2.0$}
 \label{fig3}
 \end{figure}

\subsection {Nucleation at Several Defect Sites}

Now we consider  a  more realistic case of several defect sites
where the nucleation can occur. Here, we assume a random
distribution of defects.  In the present simulation, we choose
${j_{max}}=16$ (nearly $0.1\%$ of the total number of sites and N
=128) and consider $\sigma_0(\vec{r_j})$  to be uniformly
distributed in the interval $[-0.3,0.3]$. All other parameters are
same as that for the single-defect case.  Initially, the system is
in a homogeneous state with $\epsilon(\vec{r},0)$ uniformly
distributed in the interval $[-0.005,0.005]$. At $ t=0$, we turn
on the stress-field $\sigma(\vec{r})$. In Fig. 3, we show the
evolution of the system at specifically chosen instants of time.
As in the case of single- site nucleation, around $t=15$,
nucleation of the product phase is seen to occur at several sites.
By $t=17.5$, these nuclei grow into twinned lenticular shape.
Several additional nuclei emerge at finite distance from these
original domains. These new sites at which the product phase
nucleates  most often coincides with the pre-existing defect
sites. However, occasionally, nucleation does occur at sites where
there were no defects due to stress accumulation arising from the
long-range term, as in the case of single-site nucleation. As can
be seen from the snap shot at $t=15$, there is a rapid growth of
the product phase along $[11]$ and $[1\bar 1]$ directions forming
a cris-cross pattern of the martensite domains.  We find that
there is very little growth beyond $t=25$ and by $t=50$ the growth
practically stops.  A comparison of this final configuration with
the corresponding single-site nucleation shows that the final
morphologies  are very similar implying that the morphology
evolution is independent of the original defect ( stress-field)
configuration. As for the morphology, we can see thin needle-like
structures emerging from larger domains.  Figure 2 also shows  the
time evolution of the area fraction $\phi$ for multi-defect case
($\bullet$) which is very similar to the single-defect case, which
again emphasizes the fact that the time evolution is not sensitive
to  initial defect configuration.  All the adaptive domain
adjustments take place beyond this time regime. Thereafter, the
area fraction saturates.

We briefly remark on the influence of the two parameters $\beta$
and $\gamma$  on the morphology. The quantity $\gamma$ represents
the strength of dissipation and $\beta$ the strength of the
transformation induced long-range interaction. We find that the
lateral width of the arms for smaller $\beta$ is larger than that
corresponding to larger values of $\beta$  for a fixed value of
$\gamma$. The area fraction is higher for smaller $\beta$  which
can be attributed to the fact that a lower value of $\beta$
corresponds to a  lower interaction between the transformed
regions and hence,lower energy cost to grow transverse to the
direction of propagation. ( Note that  $\beta = 0$ corresponds to
symmetric growth.)  Now consider the influence of the dissipative
term on the morphology. Our finding is that  the twin width is
larger for larger values of $\gamma$. This result can be
understood by noting that the overdamped case ($\gamma >> 1$)
corresponds to growth as a single variant \cite{Bales}. With the
inertial effects, growth as a single variant is prohibited by high
kinetic energy cost. In view of this, for large values of
$\gamma$, the dissipative term dominates and the system grows as a
single variant for a larger distance than that for the low damping
case.

\section{Thermal Cycling and Hysteresis}

\subsection{Morphological Features}

One key feature of a martensitic transformation is the hysteresis
observed when the system is subjected to thermal cycling. Here,
the system is cooled at a specific rate from the high temperature
phase to the low temperature phase and then subsequently heated
back to the high temperature phase.  We have performed `continuous
cooling' and 'heating' computer simulations where we change $\tau$
at a constant rate: the interval $\tau = 40 $ to -80 is cooled in
1000 time steps. We have monitored both the morphology and the
area fraction of the transformed phase for single-defect
nucleation as well as multi-defect nucleation case with several
system sizes ( N=128,256 ). We use the same initial conditions for
both these cases as that used for a single quench situation. The
initial condition for the reverse transformation is the final
configuration obtained during the cooling run.

In  Fig. 4, we have shown the variation of the area fraction
$\phi_A$ with $\tau$ for the heating and the cooling runs for the
single  defect case ($\bullet$) and multi defect case ($\circ$)
for $N = 128$. In the cooling run, for the multi-defect case, the
transformation starts around $\tau \sim - 2.0$ showing a rapid
increase in $\phi_A$ ( $\sim 30\%$). Thereafter, there is a nearly
linear increase up to $90\%$ at which the growth rate tapers off.
The system gets fully transformed at $\tau_{mf} \sim -60$. In the
heating run, the reverse transformation does not start till
$\tau_{as} \sim -22.0$ and thereafter, $\phi_A$ decreases almost
linearly  till the transformation is nearly complete around $\tau
\sim 18$. As can be seen from the figure (Fig. 4), the difference
between the hysteresis cycles corresponding to the multi- site
defect nucleation and single-defect nucleation cases is small.

\begin{figure}
\includegraphics[height=4cm]{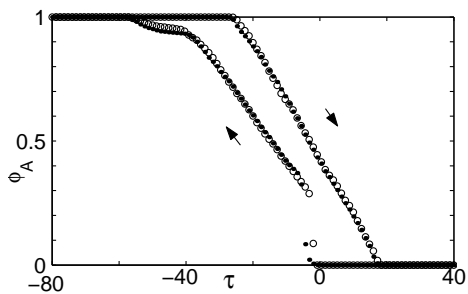}

  \caption{Area fraction of the transformed phase $\phi_A$ as a function
of $\tau$ for both single
  defect ($\circ$)  and multi-defect ($\bullet$)  nucleation cases.
  The parameter values are $\beta=50, \gamma=4$ and $N =128$.
  The defect density in multi-defect is $0.1\%$ of sites. }
  \label{fig4}
\end{figure}

\begin{figure}
\includegraphics[height=4.5cm]{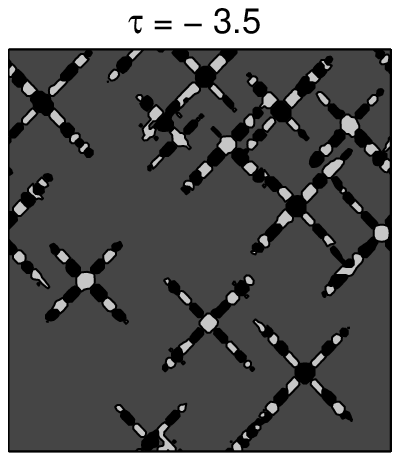}
\includegraphics[height=4.5cm]{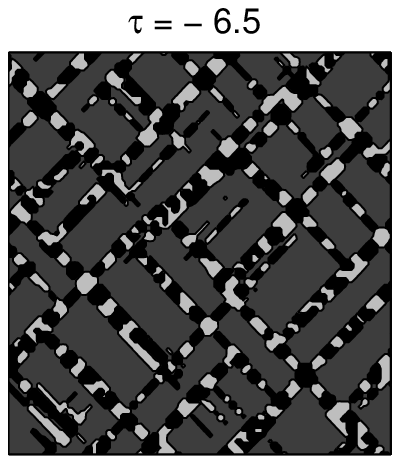}
\includegraphics[height=4.5cm]{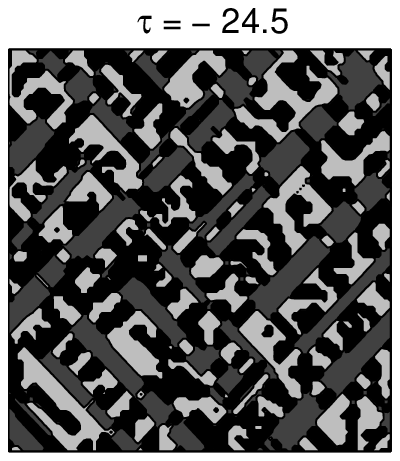}
\includegraphics[height=4.5cm]{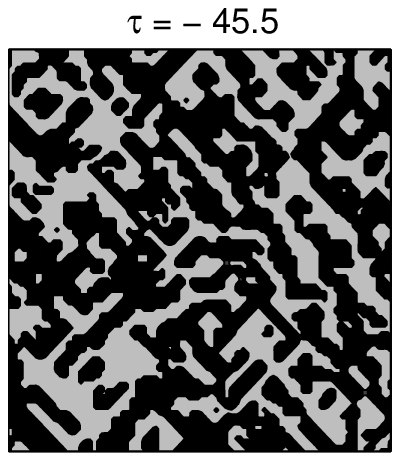}
\includegraphics[height=4.5cm]{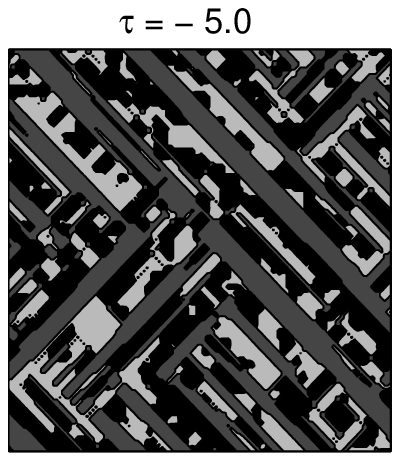}
\includegraphics[height=4.5cm]{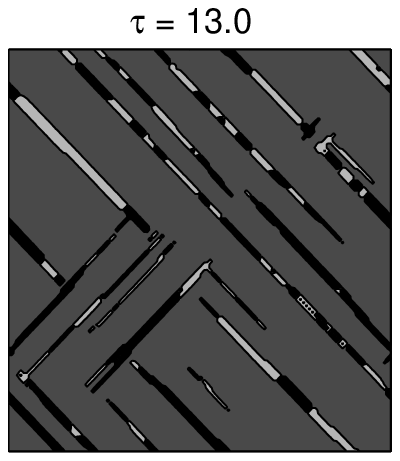}
  \caption{ Morphological evolution during a cooling
  and heating cycle. The parameter values are $\beta=50, \gamma=4$ and $N =128$
with a   defect density $0.1\%$ of the sites. }
  \label{fig5}
\end{figure}

We have followed the morphological evolution of the martensitic
domains both for the single and multi-defect cases.  Here, we
shall only discuss the more realistic multi-defect case. Figure 5
shows the snap shots of the pattern at specifically chosen values
of $\tau$ as the system is taken through a cycle. In the snapshot
corresponding to $\tau = -3.5$, one can see the nucleation of
martensitic domains at multiple locations. As the system is
further `undercooled' to lower values of $\tau$, not only does
these twins propagate in the $[11]$ and $[1\bar1]$ directions, the
thickness of the martensite domains also increases. This can be
seen from the snapshots corresponding to $\tau = -6.5$ and $\tau =
-24.5$. As the system is further undercooled, the twin width
further increases with the last regions transforming around $\tau
= -59$. A late stage snap shot at $\tau = -45.5$ corresponding to
$\phi \sim 90\%$ is shown. For the heating cycle, the final
configuration of the cooling run is taken as the initial
configuration. The austenite phase appears around $\tau \sim -22$.
As $\tau $ is further increased, the martensite phase can be seen
to be gradually disappear in the snapshots corresponding to $\tau
= -5$ and $13$.  Further, one can see that the overall morphology
in the final stages of the heating run is significantly different
from the initial stages of the cooling run. Thus, in our model,
there is no long term memory effects, though there is short term
memory as will be shown later. Finally, by $\tau = 18.0$, the
martensite phase disappears completely.

\subsection{Power Law Statistics During Thermal Cycling}

From Fig. 4, it appears that changes in $\phi_A$ are smooth on the
scale shown in the figure. In reality,  on a finer scale, the
changes in $\phi_A$ are actually jerky. In fact, in experiments,
thermal cycling is accompanied by the emission of acoustic energy
in the form of bursts, a feature that reflects the jerky nature of
the transformation. In the model, as mentioned earlier, the energy
of acoustic signals is captured by  the rate of energy dissipated
given by $R(t) = -dE/dt $. We have calculated $R(t)$ during the
heating and cooling runs. Figure 6 shows $R(t) $ as a function of
$\tau $ with the inset showing the enlarged section of the peak. The
figure clearly shows that the rate of energy release occurs in
bursts consistent  with acoustic emission studies \cite{Vives94}
during thermal cycling.

Since, in experiments  one finds that  the AE signals show a power
law statistics, we have investigated the distributions of the
amplitudes of the AE signals and their durations. Denoting the
amplitude of $R(t)$ by  $R_A$, we find that the distribution
$D(R_A)$ of $R_A$ has a tendency to approach a power law, ie.,
$D(R_A)\sim A^{-\alpha_R}$ with an exponent $\alpha_R$. Figure
\ref{fig7} shows a log-log plot of $D(R_A)$ as a function of
$R_A$, for both the single-site nucleation case ($\bullet$) and
the multi-defect case ($\circ$). From the figure, it is clear that
both these cases exhibit the same exponent value $\alpha_R \sim
2.5$ over three orders in $D (R_A)$. In experiments, one also
finds that duration of these bursts also obey a power law
statistics. To verify this, we have also plotted the distribution
$D(\Delta t)$ of the durations $\Delta t$ of energy bursts for
both the single and multi defect cases. We find that $D(\Delta t)
\sim \Delta t ^{-\tau_R}$ with an exponent value $\tau_R \approx
3.2$, although, the scaling regime is not as impressive as for
$R_A$. ( Here, we remark that typically the scaling regime for the
durations of the events is much smaller than that for the
amplitudes  even in models of SOC \cite{Bak,Jensen}.) We have also
calculated the conditional average $<R_A>_c$ for a given value of
$\Delta t$ \cite{Vives1}. This is expected to obey a power law
given by $<R_A>_c \sim \Delta t^{x_R}$. The value we get is about
$x \approx 1.36$ for both these single and multi defect nucleation
cases. Using these values, we find that the scaling relation $
\alpha = x ( \beta - 1) + 1$ is satisfied quite well. We have also
carried out a similar analysis on $R(t)$ for the heating run.
Even though, the changes in $R(t)$ occurs in bursts, we find the
scatter is considerably more than for the cooling run. We will
comment on this aspect later.

A comment is in order regarding the exponents. In experiments one
actually plots the distribution  for the amplitude of the AE
signals,while in our model acoustic energy plays a natural role.
Thus, while comparing the exponents values, one needs to use $R_A
\sim A_{AE}^2$, where $A_{AE}$ is the amplitude of the AE signal.
Using the relation between the two joint probability distributions
$D(R_A,\Delta t) \propto D(A_{AE}, \Delta t)/A_{AE}$, one easily
finds that $\alpha_R = (\alpha_{AE} + 1)/2$ with the other two
exponent values remaining unchanged. Using the experimental values
\cite{Vives94} of $\alpha_{AE} \approx 3.8$, ($\tau_{AE} \sim 3.8$
and  $x_{AE} \sim 1)$, we see that the calculated value of
$\alpha_R \approx 2.4$. Considering the fact that  our model is
two dimensional, we see that the agreement of the exponent values
is reasonable. Here, it should be pointed out that even the number
of martensite variants in 3-D are more than in 2-D. Thus,  it
would be unrealistic to expect that the mechanisms in 3-D can be
accounted for in 2-D.
\begin{figure}
\includegraphics[height=5cm]{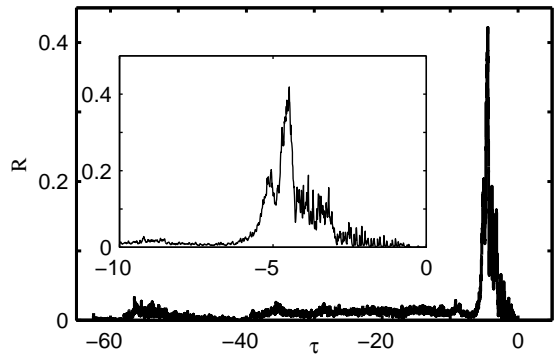}
\caption{R($\tau$) as a function of $\tau$ with inset showing
enlarged section of the peak during cooling. The parameter values
are  $\beta=50, \gamma=4$ and $N =256$.  The defect density in
multi-defect is $1\%$ of sites.
}
 \label{fig6}
\end{figure}

\begin{figure}
\includegraphics[height=6cm]{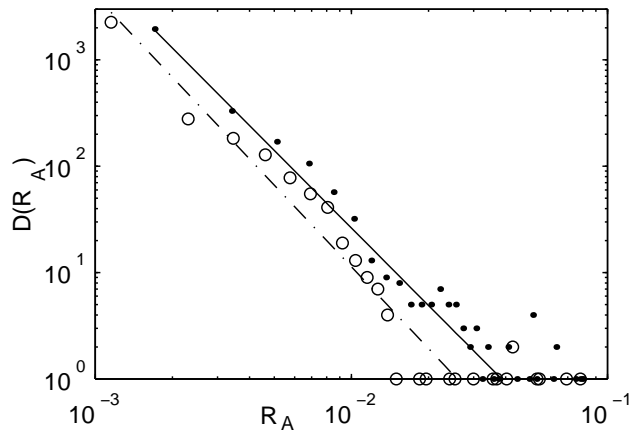}
\caption{log-log plot of $D(R_A)$ as a function of $R_A$. The
parameter values are  $\beta=50, \gamma=4$ and $N =256$.  The
defect density in multi-defect is $1\%$ of sites. [$\circ$
corresponds to single-defect case and $\bullet $ corresponds to
multi-defect case.] } \label{fig7}
\end{figure}

\begin{figure}
\includegraphics[height=5cm]{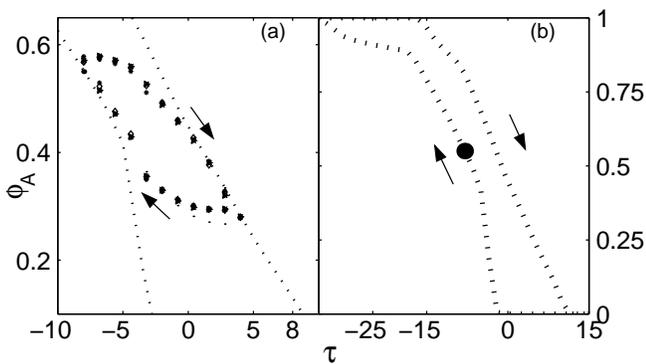}
\caption{(a).Area fraction $\phi$ during thermal cycling in the interval
 $\tau = -8$ to $ 4 $ for cycles 1 ($\circ$),2 ($\textasteriskcentered$),
3 ($\bullet$),4 ($\triangleright $).
  (b)Hysteresis for the full cycle from the austenite to the
 martensite phase and back. ($\bullet $ ) is the starting point of the small
 thermal cycles. } \label{fig8}
\end{figure}

We now explain the origin of the power law statistics in the
model. This can be traced to the fact that we have included
important ingredients of SOC dynamics, namely, the threshold
dynamics, dissipation, the generation of large number of
metastable states during the transformation, and  a relaxation
mechanism for the stored energy. The relaxation of the stored
energy occurs at very fast  time scales comparable to time scale
of the speed of sound as can be seen from the fact that the basic
variables are the displacement fields. ( We note that this is  the
fastest time scale.)  Indeed, from our simulations  we find that
the interface movement occurs at time scales of a few units of
(scaled) time as can be inferred by rapid increase in the area
fraction of the transformed phase, typically of the order of a few
units of scaled time. (See Fig. \ref{fig2}.) Compared to this the
driving force generated by thermal cycling increases with
temperature slowly   which is one of the characteristic features
of SOC dynamics.

 Another important feature of the model, as also
that of SOC dynamics,  is the creation of large number metastable
states during cooling or heating runs. This  is a direct
consequence of an interplay between the local free energy ( free
energy barrier) and the long-range interaction between the
transformed domains as can be seen from the following reasoning.
We note that the value of the long-range term at any spatial
location is the result of the superposition of the contributions
arising from the spatial distribution of the already transformed
domains. As a consequence, the free energy surface at any given
time is a complex terrain of local barriers ( metastable states).
It must be noted that these local thresholds are {\it self
generated} ( transformation induced). At a given time, these local
thresholds must be overcome by the increase in the driving force
arising from the slow cooling (or heating). We note that once a
local barrier is overcome, part of the driving force goes in
creating a new twin and the rest is dissipated in the form of
burst of energy  due to the advancing one or more interfaces. The
fact that long-range interaction is at the root of creating the
local thresholds is {\it further supported by the fact that we
find a power law distributions even in the single- site nucleation
case}. (See $\bullet$ in Fig. \ref{fig7}.) The presence of defect
sites only serves to trigger the initial nucleation process. This
must be contrasted with disorder based Ising models\cite{Sethna}
which also produce power law statistics for avalanches and field
induced hysteresis. However, as mentioned earlier, Vives {\it et
al} have verified that in martensite transformation, it is the
dynamical disorder ( transformation induced) that is at the root
of the avalanches. This exactly what is well captured by the
present model.

\section{Shape Memory Effect}

\subsection{Correlated behavior of AE signals }

As mentioned in the introduction, one experimental result (known
for some time) is the highly correlated behavior of AE signals
when the system is subjected to thermal cycling in a small
temperature interval.  To verify if this result can be captured by
our model,  consider the system being cooled from the austenite
phase to a point in the martensite phase where a desired amount of
martensite phase has developed. Starting from an appropriate point
( shown as $\bullet$ in Fig. \ref{fig8}b, here chosen to be $\phi
\sim 0.58$ in the full thermal cycle), we subject the system to
repeated thermal cycling in a small temperature range
$\tau_{min}=-8 $ and $\tau_{max}= 4$. As in the case of full
thermal cycling, for the small heating cycle also, the final
configuration attained at $\tau_{max} = 4$, is taken as the
initial configuration for the cooling run. Calculations have been
performed for a range of parameter values of $ 50 \le \beta \le
10$ and $5 \le \gamma \le 1$. For the present calculation, we have
used $\beta=35$ and $\gamma=4$. The value of $\beta$ used here is
higher than that reported in our earlier paper \cite{Kala}. We shall discuss
the influence of increasing $\beta$ soon. Other parameter values
are $ \Lambda = 0.2,\zeta = 1$ and $N=128$. During the first few
cycles, the loops in the area fraction $\phi$ verses $\tau $ drift
slightly, but stabilize after a first few cycles, here, after the
sixth cycle. The first few cycles play the role of the {\it training}
period known in experiments. After these first few cycles the
system eventually circulates in the same set of configurations as
we will show. ( The number of training cycles in the model is only few,
however in experiments, the training period is typically a few
cycles for $CuAl Zn$ \cite{Lovey}, but could be much larger in
some other alloys.)

\begin{figure}
\includegraphics[height=4cm,width=8.5cm]{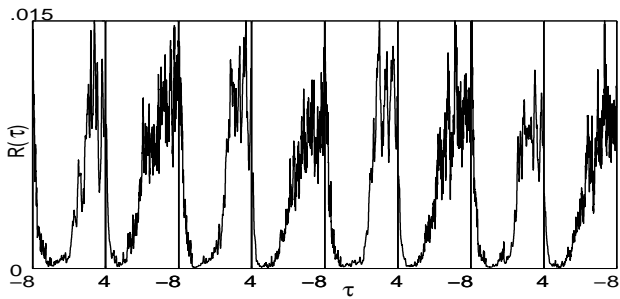}
\caption{Repetitive nature of $ R(\tau) $ for cycles 7 to 10.}
 \label{fig9}
\end{figure}

During the training cycles,  the energy dissipated $R(t)$ evolves
continuously, stabilizing only after the training period. A plot
of $R(t)$ for several forward and reverse cycles  ( seventh  to
tenth ) after stabilization is shown in Fig. \ref{fig9}. It is
clear that the energy bursts (which mimic the AE signals), as in
experiments \cite{Amen,Picor}, exhibit a near repetitive pattern
in time (temperature) during successive heating and cooling parts
of the cycles. As can be seen the bursts are much more noisy when
compared to slightly smaller value of $\beta$ used in our earlier
study\cite{Kala}. Further, we find that increasing $\beta$
requires more cycles to stabilize ( for instance, the cycles
stabilize after the fourth for $\beta =25$).

\subsection{Connection to shape memory effect}

\begin{figure}
\includegraphics[height=4.5cm]{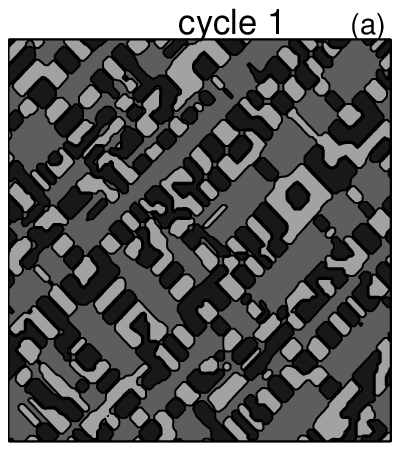}
\includegraphics[height=4.5cm]{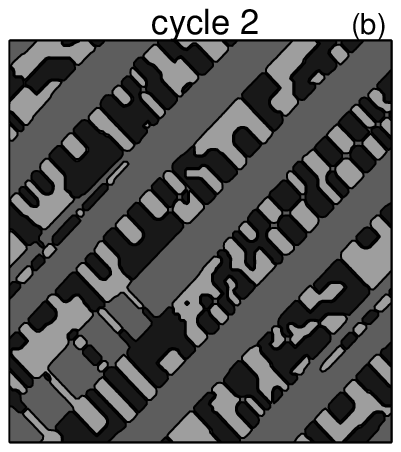}
\includegraphics[height=4.5cm]{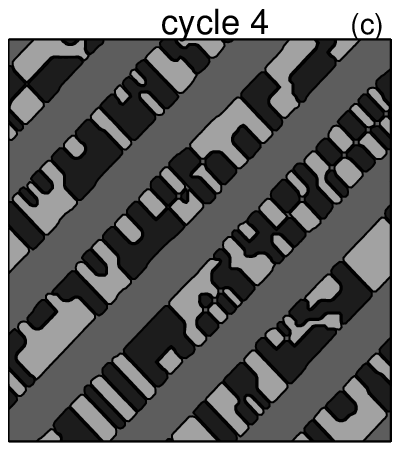}
\includegraphics[height=4.5cm]{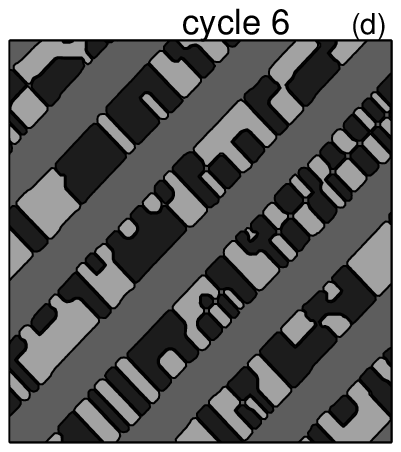}
\caption{Morphology of the initial configurations at $\tau = -8.0 $ for the
first, second, fourth and the sixth thermal cycles.} \label{fig10}
\end{figure}

In order to establish a correspondence between $R(t)$ and the
changes in the spatial configuration of the martensite domains, we
have simultaneously monitored the morphology over all the cycles.
We find that the morphology changes drastically during the first
few cycles even though the macroscopic state of system in terms of
$\phi - \tau$ returns to nearly the same point  at the end of each
cycle. The domain configurations at the beginning of  first few
cycles are shown in Fig. \ref{fig10}.   Figure \ref{fig10} a is
the starting morphology for the first cycle that has been obtained
by slowly cooling from the austenite phase. Figure \ref{fig10}d
is that obtained at the end of sixth cycle. As can be seen, most
changes occur during the first few cycles with large number of
changes occurring in the first cycle itself.  We find that during
the first cycle itself, most of the curved twin interfaces in the
initial configuration ( Fig. \ref{fig10}a) are rendered straight
and several small twins coalesce to form a single variant of the
same type. One can also notice that some regions of the austenite
phase separating the martensite domains are washed out. Subsequent
cycles also have the same effect, but less effective. The
morphology stabilizes even though some curved interfaces with
small regions of martensite phases remain ( see the top left edge
in Fig. \ref{fig10}d ). After the sixth cycles very little changes
could be detected.

\begin{figure}
\includegraphics[height=4.5cm]{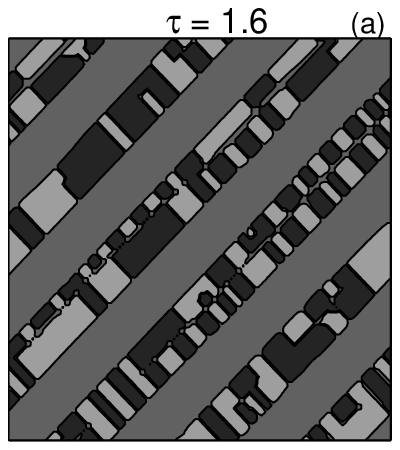}
\includegraphics[height=4.5cm]{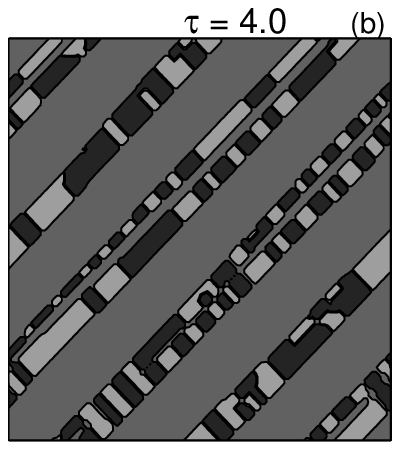}
\includegraphics[height=4.5cm]{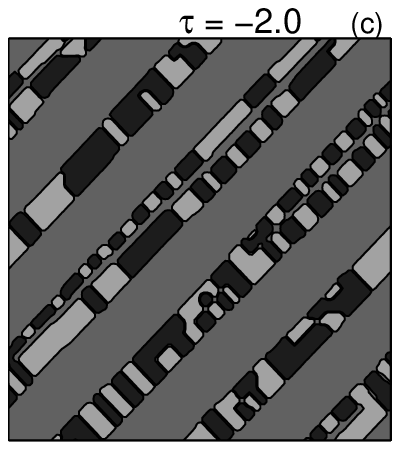}
\includegraphics[height=4.5cm]{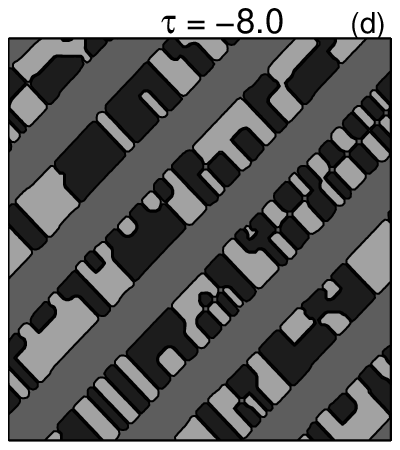}
\caption{Sequential morphological snapshots for $\tau = 1.6 (a), 4.0 (b), -2.0 (c), -8.0(d)$ during the seventh cycle. The initial configuration for the
 cycle is that shown in Fig.10(d).}
\label{fig11}
\end{figure}

Now consider the changes in the morphology during a single cycle
in the region where $R(t)$ is repetitive.   The snapshots of the
morphology during one such stabilized cooling and heating cycle,
the seventh one, (starting from the initial configuration shown in
Fig. \ref{fig10}d ) at selected intervals is displayed in Fig.
\ref{fig11}. It is clear that during heating, the martensite
domains shrink, opening up the austenite phase and some martensite
domains even disappear. However, during cooling these domains
reappear and the eventual morphology at the end of the cycle is
practically recovered on returning to the starting point on the
$(\phi,\tau)$ diagram. As can be seen, the final configuration
obtained during the seventh cycle, {\it Fig. \ref{fig11} d can be
seen to be practically the same as Fig. \ref{fig10} d which is the
initial configuration for the seventh cycle.}   These observations
are consistent with that observed in experiments \cite{Lovey}.

Now, we attempt to provide a physical explanation for   the
repetitive nature of the energy bursts and the reversal of the
morphology under thermal cycling. To understand this, we need to
understand the role played by  the training cycles. Indeed, the
repetitive pattern of the energy bursts during successive cycles
(after the training period) is an indication that the system
traverses through the same set of metastable states. From Fig.
\ref{fig10}a, we note that the initial configuration for the first
cycle has a large number of small domains compared to the
configuration obtained after  the sixth cycle shown in Fig.
\ref{fig10} d. In addition, the twin interfaces of Fig.
\ref{fig10}a are rough. ( The interface is considerably more
curved for small $\beta$ values. Compare Fig. 3 of Ref.
\cite{Kala}. For the present case, it is relatively straight.)
Such configurations are generally expected to have higher energy
compared to straighter ones. Thus, the initial configuration used
for cycling (Fig. \ref{fig10}a) corresponds {\it only} to a local
shallow minimum. During the first few cycles, the free energy
landscape is so modified that it smoothen's out the energy
barriers corresponding to  large number of twin interfaces in Fig.
\ref{fig10}a) with very little change in the area fraction. As in
the case of power law distribution, a crucial role in smoothening
process is actually played by the long-range interaction term, as
the growth (shrinkage) of a martensite domain is influenced  by
the configuration of rest of the domains. To verify this, we have
computed the free energy $F_{LR}$ arising from the long-range
interaction between the domains and find that  it actually
becomes more negative with successive cycles saturating after
first few cycles. This additional contribution  leads to a
reduction in the local free energy, $F_L$, as well. The net effect
is to create {\it a deeper set of metastable states for the system
to circulate for the stabilized cycles}. Within one such
stabilized cycles, say seventh, the starting configuration (Fig.
\ref{fig10}d) has the lowest free energy reaching a maximum at the
end of a heating cycle, ie., at $\tau = 4.0$, Fig. \ref{fig11}b.
Thereafter it decreases during cooling.  Note that the morphology
at $\tau = 4.0$  is significantly different from the starting
morphology.  The increasing or decreasing  width of the martensite
domains as we decrease or increase the temperature is surprisingly
similar to that observed in experiments. ( See Fig. 12 of
Ref.\cite{Lovey}.)

\begin{figure}
\includegraphics[height=4.5cm]{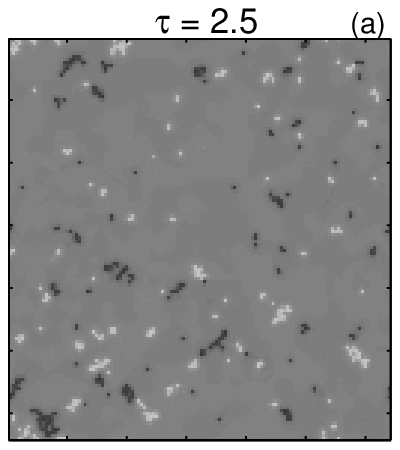}
\includegraphics[height=4.5cm]{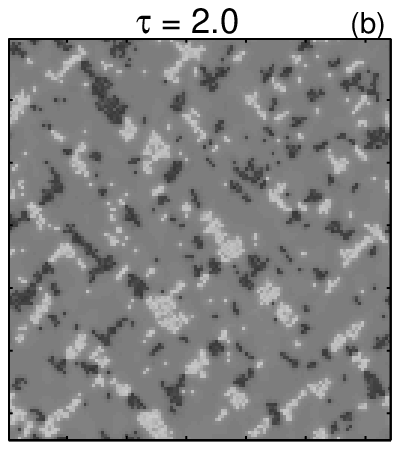}
\includegraphics[height=4.5cm]{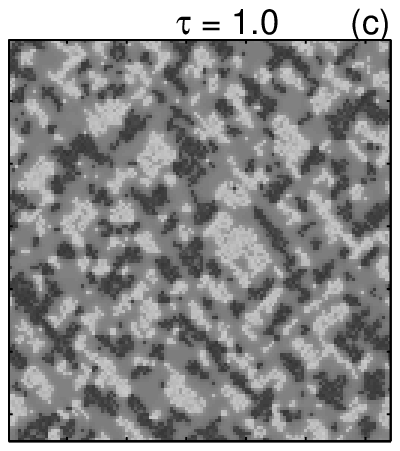}
\includegraphics[height=4.5cm]{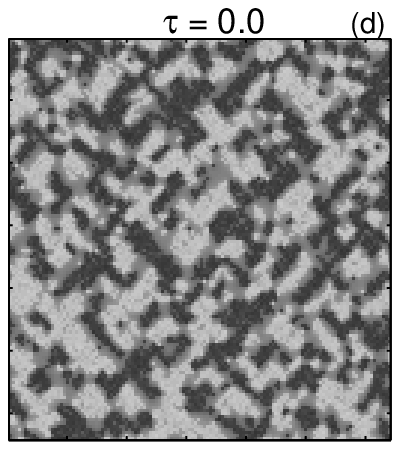}
\caption{Development of Tweed like pattern} \label{fig12}
\end{figure}

\section{Pretransitional effects }

Now we consider the possibility of recovering pretransitional
effects in our model. As mentioned in the introduction, the
mechanism attributed  by Kartha {\it et al} is the dependence of
the elastic constant on local disorder in
composition.   It is clear that once the basic mechanism is
included, the model should lead to results similar to that  in Kartha
{\it et al}. We shall  adopt the same idea and assume that the
transition temperature depends on local compositional fluctuations
through
\begin{equation}
T_0(\vec r) = T_0(\bar c) - A \delta c (\vec r),
\end{equation}
where $A$ is the relative strength of coupling to  compositional
fluctuations $\delta c$, assumed here to be randomly distributed.
In scaled variable that we use, this leads to $\tau $ being
replaced by $ \tau= \tau(\bar c) + a \delta c$, where $\bar c$ is
the average concentration and $a$ is a scaled variable.

In our simulations, we assume that the initial concentration
fluctuations, considered as frozen, are drawn from a normalized
Gaussian.   In our case, we just need to solve the equations of
motion numerically  for various values of $\tau$ and $a$. (We have varied
 $a$  from 0.0 to 1.2.) The range of values of
$\beta$ and $\gamma$ wherein we find the tweed structure is in the
region of relatively low values.( $\beta < 20$ and $\gamma<0.7$.)
Here, we report results for $a = 0.9, \beta =10$ and $ \gamma
=0.1$. A typical set of morphologies are shown in Fig.
\ref{fig12}. The pretransitional effects become noticeable even
when $\tau = 4$. For $\tau=2$ the directionality of the tweed
pattern is already evident. As we decrease the temperature, this
structure becomes more dominant. It must be mentioned here that in
our case, these patterns do not change after reaching a steady
state. Thus, we do not have the dynamic tweed structure reported
by Kartha {\it et al}. This is because, in our model the effect of
temperature goes only in the local free energy and there are no
thermal fluctuations. We also note that the pretransitional
effects are pronounced when both $\beta$ and $\gamma$ are small.
We shall comment on this later. However, we note that including
the changes in the transition temperature on local disorder  is
somewhat similar to the contribution from defects except that
internal stresses ( arising from deviations from the average
concentration) are of much smaller magnitude compared to that due
to defects. Thus, it is clear that one can mimic the present
situation through Eq. \ref{defect} by appropriate choice of
$\sigma_0$ and $\zeta_i$. In our case, the $a-\tau$
 plane, the boundary between the austenite and the tweed phases is linear in
 the range (0,0) to (1.2,4).

\section{SUMMARY AND DISCUSSIONS}

In summary, we have presented the results of a comprehensive study
of the dynamics of strain driven martensitic transformations
within the framework of a two dimensional square-to-rectangle
transition. Due to the fact that the long-range term is introduced
phenomenologically,  the model should be viewed as a model whose
primary aim is to capture the essential physics of the
transformations.  We however note that it includes all the
important contributions arising from different mechanisms in a
transparent way. In doing so, we are able to study the dynamics of
the strain driven transition that  explains  the three unusual
features of the martensitic transformation that were sought to
modelled. The first important feature that emerges from the model
is the fact that the elastic energy stored is released in the form
of bursts. (Note  that we have also established a correspondence
between the dissipative functional and the energy of AE signals.)

Second, the model provides a proper basis to explain the
power law statistics of the AE signals observed in experiments. As stated
 earlier, the power law statistics arises due to the fact
that we have included threshold dynamics, dissipation, the
generation of large number of metastable states during the
transformation, and a relaxation mechanism for the stored energy
whose time scale is much faster than the time scale of the drive
force. It is interesting to point out that both cases of single-defect and
multi-defect nucleation lead to power laws strongly suggesting that quenched
 defects do not play any role in the power law.
The model also predicts the near repetitive bursts of
energy under successive thermal cycles in a small temperature
interval as observed in experiments on acoustic emission
\cite{Lovey,Picor,Amen}. This comes as a surprise as the nature of
 correlation in the latter case is 'periodic' in contrast to the scale
free nature of correlations discussed in the power law statistics
of AE signals. More importantly,
these bursts of energy have been shown to be correlated to the
growth and shrinkage of martensite plates. The underlying cause of
the correlated nature of AE signals and reversal of the morphology
during successive cycles after the training cycles  has been
traced to the influence of the long-range interaction term. The
role of training cycles is also elucidated. During the training
period the long-range term has a tendency to smoothen out higher
energy barriers in the free energy landscape. This in turn induces
a transformation pathway along a unique set {\it of low energy
metastable configurations}. We expect that our analysis provides a good
insight into the shape memory effect which finds immense applications in a
variety of areas from mechanical actuators to bio-medical applications
\cite{Van}. To the best of authors knowledge, this is the first model which
shows near full reversal of morphology under thermal cycling. Finally,
it is not surprising that the model also reproduces tweed like features.

 A comment may be in order on the
physical interpretation of the cause of the  memory effect in our
model. Recall that during the course of the transformation (either
in single quench situation or in thermal cycling), nucleation
occurs at sites that are not necessarily the sites of quenched
defects, but arising from  constructive superposition of  long-
range stress fields of the preexisting domains. However, as
quenched defects have been represented by a Gaussian stress
profile in our model, in the same spirit, these new nucleation
sites  could in principle be interpreted as creating new defects
at these sites. (See the comment in Ref. \cite{Reche}.) We also
believe that it may be possible to model this with an additional
degree of freedom for the defect kinetics.  We expect that our
analysis provides a good insight into the shape memory effect
which finds immense applications in a variety of areas from
mechanical actuators to bio-medical applications \cite{Van}.

It is interesting to note  the similarity of morphological patterns
 with real micrographs. Patterns studied in the
context of single-defect and multi-defect nucleation show that the
eventual morphology are independent  of the the original defect
configuration. It is not clear if this is true of real systems
although experiments of Vives {\it et al} \cite{Vives94} provide
an indirect evidence. In addition, the model shows hysteresis
under thermal cycling.  Even though, the model explains several
generic features stated above, we stress that our model is not
material specific.

As mentioned the pretransitional effect is pronounced when both
$\beta$ and $\gamma$ values are small. The low  values of
$\gamma$ is physically understandable as one expects that there
would be hardly any  dissipation in the high temperature phase where
there are no martensite domains. The smallness of $\beta$ is also
understandable  physically (as  the magnitude of
fluctuations in the composition is not high). However, the value of
$\beta$ cannot be determined in terms of the elastic constants
as in the case where the kernel is derived by using the compatibility
relation \cite{Kartha95}.

Some  comments may be in order on SOC type of features obtained in the
 model. We
note here that unlike most SOC models where noise is essential
\cite{Jensen,Bak},  the model is fully dynamical in the sense that
noise has no role in the generation of power law statistics.  In
spirit, the model is closer to that by Gil and Sornette\cite{Gil}
where  an explicit threshold term is introduced in the form of a
subcritical Hopf bifurcation. However, it must be pointed out that
noise is essential in their model as well. Lastly, to the best of
the authors knowledge, this is one of the few  fully continuous
space-time dynamical (noise free) model for SOC.

\begin{acknowledgements}
  Part of this work was carried out when one of
the authors (RA) was supported by JNCASR which is gratefully
acknowledged.
\end{acknowledgements}

\appendix*

\section{}
Here we present some details of the derivation of Eqn. 15.
 To start with we evaluate ${ {\delta F_{L}} \over {\delta u_{x}} }$ ,
\begin{equation}
{ {\delta F_{L}} \over {\delta u_{x}} }=\int d\vec{r}
\bigg[\bigg({ {\partial f} \over{\partial \epsilon(\vec{r})}
}-\sigma(\vec{r})\bigg) {{\delta \epsilon(\vec{r})}\over{\delta
u_{x}(\vec{r'})}}+ D (\nabla \epsilon ) {{\delta
\nabla\epsilon(\vec{r})}\over{\delta u_{x}(\vec{r'})}} \bigg],
\end{equation}
Using the definition for $\epsilon$, the above expression  reduces
to a compact form
\begin{equation}
{ {\delta F_{L}} \over {\delta u_{x}} }
={ {\partial} \over { {\partial x}} }\bigg[
-\bigg({ {\partial f}
\over{\partial \epsilon(\vec{r})} }-\sigma(\vec{r})\bigg)
+ D {\nabla}^{2}\epsilon(\vec{r})\bigg] \nonumber
= -{ {\partial} \over { {\partial x}} }\bigg[{{\delta F_{L}}\over{\delta\epsilon(\vec{r})}}
\bigg].
\end{equation}
An analogous calculation for $\delta F_L \over \delta u_y$ holds except for
a negative sign on the left
hand side with interchange of the $ x $ with $y$.
Similarly, one can show that
\begin{equation}
{ {\delta F_{LR}} \over {\delta u_{x}} }=
{ {\partial} \over { {\partial x}} }\bigg[\epsilon(\vec{r})
\int d\vec{r'}G(\vec{r}-\vec{r'})\epsilon^{2}(\vec{r'})\bigg]
= -{ {\partial} \over { {\partial x}} }\bigg[{{\delta F_{LR}}\over{\delta\epsilon(\vec{r})}}
\bigg]
\end{equation}
and a similar equation for $u_y$.
The functional derivative ($x$ component) for the dissipation functional is
\begin{equation}
{{\delta R} \over { \delta \dot{u}_x} } =
-\gamma{{\partial}\over{\partial x}}
\bigg({{\partial}\over{\partial t}}\epsilon(\vec{r},t)\bigg),
\end{equation}
and a similar equation for ${\dot u}_y$.
One also finds
\begin{equation}
{{d}\over{dt}}\bigg({{\delta L}\over{\delta {\dot{u}_i}}}\bigg)
= \rho{{{\partial}^{2} }\over{{\partial t}^{2}}}u_i(\vec{r},t), i=x,y.
\end{equation}
Then, using these results in Eq.(13), and taking appropriate derivatives with
respect $x$ and $y$,  we get the equation for
the strain order parameter as
\begin{eqnarray}
\nonumber
\rho{{{\partial}^{2} }\over{{\partial t}^{2}}}\epsilon(\vec{r},t) & = &
{\nabla}^2\bigg[{{\partial f(\vec{r},t)}\over{\partial \epsilon(\vec{r},t)}}
 -\epsilon(\vec{r},t)\int d\vec{r'}G(\vec{r}- \vec{r'})\epsilon^{2}(\vec{r'},t)\\
 & - & \sigma(\vec {r}) - D{\nabla}^2\epsilon(\vec{r},t)
+\gamma
{{\partial }\over{\partial t}}\epsilon(\vec{r},t)\bigg].
\end{eqnarray}
Using $F=F_L+F_{NL}$,  in a compact form the above equation  can be written as

\begin{equation}
\rho{{{\partial}^{2} }\over{{\partial t}^{2}}}\epsilon(\vec{r},t)=
{\nabla}^2\bigg[{{\delta F}\over{\delta \epsilon(\vec{r},t)}}
+\gamma
{{\partial }\over{\partial t}}\epsilon(\vec{r},t)\bigg].
\end{equation}


\begin{thebibliography}{45}
\bibitem{Remark} This choice is not unique. Further, it may be possible to derive an appropriate
 kernel which couples bilinear terms in $\epsilon(r)$ and $\epsilon(r')$. However, we expect that the functional form of $ B(\vec k/k) $ would be complicated.
\end{thebibliography}
\end{document}